\documentclass[%
preprint,
showpacs,
amsmath,amssymb,
aps,
]{revtex4-1}

\usepackage[dvipdfmx]{graphicx}
\usepackage{bm}
\usepackage{setspace}

\begin{document}

\preprint{}

\title{
Phase transition of social learning collectives and ``Echo chamber''
}

\author{Shintaro Mori}
\email{mori@sci.kitasato-u.ac.jp}
\affiliation{
Department of Physics, Faculty of Science, Kitasato University \\
Kitasato 1-15-1, Sagamihara,  Kanagawa 252-0373, JAPAN
}

\author{Kazuaki Nakayama}
\email{nakayama@math.shinshu-u.ac.jp}
\affiliation{
Department of Mathematical Sciences,
Faculty of Science, Shinshu University
\\
Asahi 3-1-1, Matsumoto, Nagano 390-8621, JAPAN
}%

\author{Masato Hisakado}
\email{hisakadom@yahoo.co.jp}
\affiliation{Fintech Lab. LLC \\
Meguro, Tokyo 153-0051, JAPAN
}%



\date{\today}

\begin{abstract}
  We study a simple model for social learning agents in a restless
  multi-armed bandit.
  There are $N$ agents, and the bandit has $M$ good
  arms that change to bad with the probability $q_{c}/N$.
  If the agents do not know a good arm, they look for it by a random
  search (with the success probability $q_{I}$) or copy the information of
  other agents' good arms (with the success probability $q_{O}$) with
  probabilities $1-p$ or $p$, respectively.
  The distribution of the agents in $M$ good arms
  obeys the Yule distribution with the power-law exponent
  $1+\gamma$ in the limit $N,M\to \infty$ and
  $\gamma=1+\frac{(1-p)q_{I}}{pq_{O}}$.
  The system shows a phase transition
  at $p_{c}=\frac{q_{I}}{q_{I}+q_{o}}$.
  For $p<p_{c}\,(>p_{c})$, the variance of $N_{1}$
  per agent is finite (diverges as $\propto N^{2-\gamma}$ with $N$).
  There is a threshold value $N_{s}$ for the system size that scales as
  $\ln N_{s} \propto 1/(\gamma-1)$.
  The expected value of the number of the agents with
  a good arm $N_{1}$ increases with $p$ for $N>N_{s}$.
  For $p>p_{c}$ and  $N<N_{s}$,
  all agents tend to share only one good arm.
  If the shared arm changes to be bad, it takes long time
  for the agents to find another good one.
  $\mbox{E}(N_{1})$ decreases to zero as $p\to 1$,
  which is referred to as the ``Echo chamber.''
\end{abstract}

\maketitle


\section{\label{sec:intro}Introduction}
Recently, social physics has become an active research field,
and many studies have been devoted to understanding the
social phenomena and interacting human behaviors
\cite{Pentland:2014,Ormerod:2012,Mantegna:2008,Cont:2000,Lux:1995,Gracia:2014}.
In the process of these studies, social learning plays the key
role as a fundamental interaction
among humans \cite{Pentland:2014,Ormerod:2012}.
Here, social learning is a learning process through observing
and imitating others' behaviors \cite{Laland:2004,Kendal:2009}.
In economics, the expected return
or expected utility is maximized in the modelling of human behaviors.
However, in many cases, such estimation is difficult or impossible for humans
because of the limited capability of calculation and information
gathering. Rationality is bounded and humans cannot avoid the
uncertainty in decision making \cite{Kirman:2010,Kahneman:2003}.
In such uncertain situations,
it is rational to observe others and copy their choices.
Social learning is easy, as the cost is
lower than individual learning, where one obtains
information with one's effort \cite{Laland:2004,Rendell:2011}.
Individual learning
has the advantage that the obtained information is more accurate and
newer than the information obtained by social learning. Even so, the cost
of the former learning is generally very high, and
this cannot be compensated for by the accuracy of the obtained
information \cite{Rendell:2010}.

As social learning is a process of copying information among agents
and agents change their behaviors and make decisions based on that
information, this process plays the role of
interaction among agents. As the interaction is
strong, the statistical mechanical approach is
promising \cite{Galam:2008,Castellano:2009}. One example is
the phase transition caused by the information
cascade \cite{Hisakado:2011,Mori:2015-2}.
Information cascade is the tendency to choose the majority choice
even if one thinks the minority option is correct or optimal
\cite{Bikchandani:1992,Devenow:1996}.
In laboratory experiments, at most 63 subjects answered
two-choice questions one by one after observing the
previous subjects' choices \cite{Mori:2012,Hino:2016}.
The sequential choice processes were described as non-linear
P\'{o}lya urn process.
It was observed that as the difficulty of
the question increases,
the number of stable states changes.
The change in the number of the stable states
induces a nonequilibrium discontinuous 
phase transition in the infinite population limit \cite{Mori:2015-2}.
The performance of humans is greatly improved by social
learning, and this is called the collective intelligence effect
\cite{Austen-Smith:1996,Surowiecki:2004}.
However, it can also cause
suboptimal macroscopic behaviors where the majority choice is
wrong.

Many kinds of social animals
undertake social learning and make collective
decisions \cite{Conradt:2009,Sumpter:2009}.
Owing to the abovementioned drawback of social learning,
it is important to investigate
when, from whom, and how one copies others.
These questions
have been studied in behavioral ethology
\cite{Giraldeau:2002,Laland:2004,Kendal:2009}.
In 2010, a paper reported a tournament of social learning
agent programs in a restless multi-armed bandit(rMAB).
A multi-armed bandit is analogous to the ``one-armed bandit'' slot
machine but with multiple ``arms,'' each with a distinct payoff
\cite{Rendell:2010}.
We call an arm with a high payoff a good arm.
The term ``restless'' means that the payoffs changes randomly.
Agents maximize their payoffs by exploiting an arm, searching
for a good arm at random (individual learning), and copying
an arm exploited by other agents (social learning).
In the tournament, strategies that relied
heavily on social learning were found to be remarkably successful.
However, when all agents always adopt social learning in
searching an arm, the performance was
very bad \cite{Rendell:2010}.
They do not bring in new information, and
outdated information is prevalent among them.
This kind of suboptimal state is called an ``Echo chamber,''
in which wrong and outdated information spreads
among agents,
and the performance is severely degraded \cite{Pentland:2014}.
By using the data in the online trade market e-Toro, it was reported
that the excessive
observation of others' portfolios leads to
the deterioration of the performance \cite{Pentland:2014}.

In this paper, we study a simple model for social learning
agents in an rMAB.
Previous studies focused on the optimal social learning strategy
in a nonuniform and changing environment \cite{Rogers:1988,Enquist:2007}.
Here, we focus on the collective behavior of social learning agents
in the context of statistical physics.
The proposed model is deeply related with other models.
If one regards an arm as a color and an agent as a ball,
it is a kind of generalized P\'{o}lya urn process \cite{Pemantle:2007}.
The bad arm's color is black, and the good arms' colors are
different from other arms' colors. If an agent knows an arm,
 the color of the ball is the color of the arm.
If an randomly chosen ball's color is black, it is replaced with a
ball with another color. Contrary to the generalized P\'{o}lya urn process,
in an rMAB, all balls of a color change to bad randomly.
In social learning, the
probability that a good arm is chosen is proportional
to the number of agents who know it. It works as a preferential
attachment mechanism of evolving networks\cite{Barabasi:1999,Dorogovtsev:2000}.
The model is also a variant of the cultural neutral model \cite{Neiman:1995}.
In this context, each arm corresponds to an option, a city, a product,
a song, a fashion, an idea, etc. In the model, in each turn, a new agent
is added, and this agent will copy the choices of previous agents.
  Instead of the spontaneous change in the arms in an rMAB, a new option is added, and
  the agent will choose this new option with a certain probability.
  The cultural neutral model explains right-skew socioeconomic
  distributions and the continuous
  turnover observed empirically in the
  distributions \cite{Evans:2007,Bentley:2011}.

We organize the paper as follows. We introduce a simple
 model for social learning collectives in an rMAB in Sec.~\ref{sec:model}.
 Sec.~\ref{sec:M1} is devoted to the analysis
 for the solvable case where the rMAB has one good arm.
 In Sec.~\ref{sec:MF}, we study the model with multiple good arms using
 the mean field approximation.
 The distribution of agents among good arms obeys the Yule distribution
 in the infinite population and good arm limit.
 We also estimate a critical system
 size $N_{s}$, beyond which the system is
 in equilibrium and the performance improves with the rate of social
 learning.
 In Sec.~\ref{sec:PTE}, we show that the system shows a
 phase transition.
 We describe how the performance behaves in each phase and
 explain the nature of the ``Echo chamber.''
 Sec.~\ref{sec:Con} is devoted to the conclusions and to discussing
 future problems.

 \section{\label{sec:model} Model}
  There are $N$ agents, and the rMAB has $M$ good arms.
  We label the agents by $n \in \{1,2,\cdots,N\}$
  and the good arms by $m\in \{1,2\cdots,M\}$.
  We identify all bad arms with the arm $m=0$.
  In each turn, one agent
  is randomly chosen, and he exploits his arm if he knows a good one.
  If he does not know a good arm, he looks for it by a random
  search (individual learning) or
  copies the information of
  other agents' good arms (social learning) with
  probabilities $1-p$ or $p$, respectively.
  In the random search, we denote the probability that the agent
  successfully knows a good arm as $q_{I}$.
  As there are $M$ good arms,
  each arm is chosen with the probability $q_{I}/M$.
  In the copy process, he can know a good arm with
  the success probability $q_{O}$ if there is at least one agent who knows
  a good one.
  If there are multiple agents
  who know a good arm, one of them is chosen randomly.
  The probability that arm $m$ is chosen is
  proportional to the number of the agents who know it.
  After that, good arms change to bad
  with the probability $q_{c}/N$ independently of each other.
  If a good arm $m$ changes to bad, the agents who
  who know it are forced to forget it and
  to know the bad arm $m=0$.
  We define one Monte Carlo step (MCS) or one round as $N$ turns.
 The probability that an arm does not change during one MCS (round) is
 $(1-q_{C}/N)^{N}\simeq e^{-q_{C}}$.

 In the agent tournament \cite{Rendell:2010},
 agents copy the arm that
 is exploited in the previous round. In each round, all $N$ agents
 perform their actions. In this case, we can estimate 
 $q_{O}$ as $q_{O}\simeq  (1-q_{c}/N)^{2N}\simeq e^{-2 q_{C}}$.
 The reasoning for this estimation is that the copied agents
 exploited the arm $N$ turns before. The information of
 the exploited arm
 was assumed to be updated $N$ turns
 before the exploit. 
 Since the distribution of the payoff obeys an exponential
 distribution, $q_{I}$ is small.
 In this paper, we assume $q_{I}<q_{O}$.

 We denote the number of agents who know arm $m$ after
 $t$ as $K_{m}(t)$.
 As we are interested in the performance of the agents' system,
 we also introduce $N_{1}(t)\equiv \sum_{m=1}^{M}K_{m}(t)$, which is the
 number of agents who know a good arm after $t$.
 The number of agents who do not know a good arm is denoted
 as $N_{0}(t)$, and $N_{0}(t)=N-N_{1}(t)=K_{0}(t)$ holds.

 In turn $t$, agent $n$ is picked randomly among $N$ agents.
 If this agent knows a good arm, the agent exploits it to obtain
 a payoff. Otherwise,
 the agent
 copies with the probability $p$ and searches randomly with
 the probability $1-p$.
 If $N_{1}=0$, the agent cannot know a good arm by copying.
In terms of $\vec{K}=(K_{0},\cdots,K_{M})$,
the probability for the change $\Delta K_{m}=1$ by an
agent action with the condition
$\vec{K}=\vec{k}$
is written as
\begin{equation}
p_{m}(\vec{k})
\equiv \mbox{Pr}(\Delta K_{m}= 1)=\frac{N_{0}}{N}
\cdot \left(A\cdot (1-\delta_{N_{1},0})\cdot
\frac{k_{m}}{N_{1}}+B\cdot \frac{1}{M}\right).
\end{equation}
Here, $A$ and $B$ are defined as $A\equiv pq_{O},B\equiv (1-p)q_{I}$.
We write the probability that there is no change in $\vec{K}$
by an agent action
as $p_{NC}(\vec{k})\equiv 1-\sum_{m=1}^{M}p_{m}(\vec{k})$.
We summarize the change in $\vec{K}$ by the transfer matrix $T^{A}$ as
\[
T^{A}(\vec{k}'|\vec{k})\equiv \Pr(\vec{K}=\vec{k}\to \vec{k}')=
\left(p_{NC}(\vec{k})\prod_{m=1}^{M}\delta_{k_{m},k_{m}'}+
\sum_{m=1}^{M}p_{m}(\vec{k})\delta_{k_{m}',k_{m}+1}
\prod_{l\neq m}
\delta_{k_{l}',k_{l}}
\right).
\]
After an agent action, all good arms change to be bad with the probability
$q_{c}/N$ independently.
If arm $m$ changes to bad, $K_{m}$ becomes zero.
We write the change in $\vec{K}$ using the transfer matrix $T^{C}$ as
\[
T^{C}(\vec{k}'|\vec{k})\equiv \Pr(\vec{K}=\vec{k}\to \vec{k}')=
\prod_{m=1}^{M}
\left((1-\frac{q_{C}}{N})\cdot \delta_{k_{m}',k_{m}}+\frac{q_{C}}{N}\cdot
\delta_{k_{m}',0}
\right).
\]

We write the probability function for
$\vec{K}(t)=\vec{k}$ as $P(\vec{k},t)$.
\[
P(\vec{k}|t)\equiv \mbox{Pr}(\vec{K}(t)=\vec{k}).
\]
The Chapman--Kolmogorov equation for $P(\vec{k}|t)$ is
\begin{equation}
P(\vec{k}|t+1)=\sum_{\vec{k}'}\sum_{\vec{k}''}
T^{C}(\vec{k}|\vec{k}')T^{A}(\vec{k}'|\vec{k}'')P(\vec{k}''|t).
\end{equation}
Initially, no agent knows good
arms, and $\vec{K}(0)=(N,0,\cdots,0)$.
The initial conditions for
$P(\vec{k}|0)$ are $P((N,0,\cdots,0)|0)=1$ and
$P(\vec{k}|0)=0$ for $\vec{k}\neq (N,0,\cdots,0)$.
If $p=1$ or $q_{I}=0$,
no agent can know a good arm forever.
We assume that $p<1$ and $q_{I}>0$ to avoid such trivial results.

If $N$ is sufficiently large, the equilibrium
between $N_{1}(t)$ and $N_{0}(t)$ should be realized in the stationary state.
We denote their equilibrium values
as $N_{1}\equiv \lim_{t\to \infty} N_{1}(t)$ and
$N_{0}\equiv \lim_{t\to \infty} N_{0}(t)$, respectively.
The equilibrium condition between
$N_{0}$ and $N_{1}$ is given as
\begin{equation}
\frac{N_{0}}{N}\cdot (A+B)=N_{1}\cdot \frac{q_{C}}{N}  \label{eq:Eq}.
\end{equation}
The left-hand side shows the expected number of agents
who come to know a good arm, which is given by the product
of the probability that a randomly chosen agent knows a bad arm,
$N_{0}/N$, and
the probability that the agent succeeds in knowing a good one, $A+B$.
If $N$ is large, the number of agents who know a good arm
is always positive, and the probability for successful copying is
$q_{O}$.
The right-hand side shows the expected number of
agents whose arm changes from good to bad.
We solve the condition with the constraint $N_{0}+N_{1}=N$, and we have
\begin{equation}
  \frac{N_{1}}{N}=\frac{A+B}{q_{C}+A+B}
  =\frac{pq_{O}+(1-p)q_{I}}{q_{C}+pq_{O}+(1-p)q_{I}}.
 \label{eq:N1}
\end{equation}
As we assume $q_{I}<q_{O}$, $N_{1}/N$ becomes
an increasing function of $p$.
In the limit $p\to 1$, $N_{1}/N$ converges to $q_{O}/(q_{O}+q_{C})$.
As will be shown later, if $p$ is larger than
a threshold value $p_{c}$, the system size $N$
required for maintaining the equilibrium
becomes extremely large.
If $N$ is smaller than the required value, $N_{1}(t)$ shows
an oscillatory behavior, as 
 the system is trapped in the
state $N_{0}=N$ for a long time.
The expected number of $N_{1}/N$ decreases with $p$.

\section{\label{sec:M1} Exact Solvable $M=1$ case}
When there is only one good arm,
the state of the system is described by
$K_{1}=N_{1}=n\in \{1,2,\cdots,N\}$.
In this case, we can obtain
the stationary probability function
$P(n)=\lim_{t\to\infty}\mbox{Pr}(N_{1}(t)=n)$ and
estimate several quantities.
Here, we explain some important results.
The details of the derivations are given in Appendix A.

The expectation value of $N_{1}/N$ is evaluated as
\begin{eqnarray}
\mbox{E}(N_{1}/N)&=&\frac{1+a/N}{a+1}\cdot (1-P(0)),\label{eq:N1_M1} \\
P(0)&=&\frac{q_{C}}{q_{C}+(N-q_{C})B}. \nonumber
\end{eqnarray}
In the limit $N\to \infty$, $P(0)\to 0$, and we obtain
\[
\lim_{N\to \infty}\mbox{E}(N_{1}/N)=
\frac{1}{a+1}=\frac{A+B}{q_{C}+A+B}.
\]
The result is in accordance with the
result from Eq.~(\ref{eq:N1}).

\begin{figure}[htbp]
\begin{center}
\includegraphics[width=10cm]{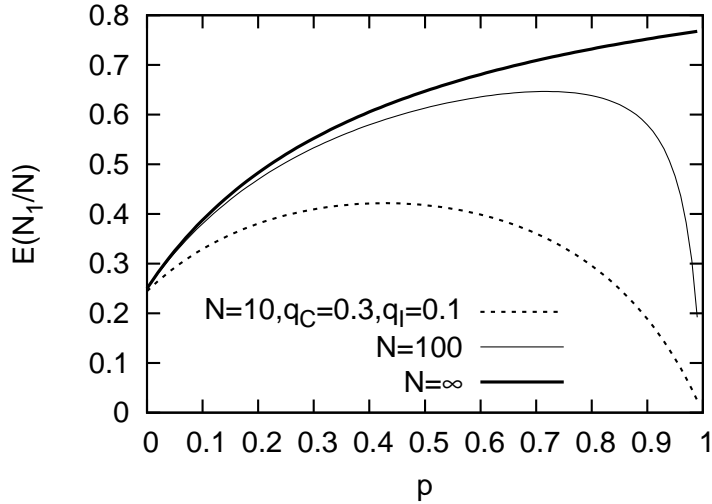}
\end{center}
\caption{\label{fig:Ez_M1}
Plots of
E$(N_{1})/N$ vs. $p$ for $M=1$. We plot Eq.~(\ref{eq:N1_M1})
for the plot $N<\infty$ and
Eq.~(\ref{eq:N1}) for the plot $N=\infty$.
$q_{C}=0.3,q_{I}=0.1,q_{O}=1$, and
$N\in \{10,10^{2},10^{4},\infty\}$.
}
\end{figure}	

Figure \ref{fig:Ez_M1} shows a plot of E$(N_{1}/N)$ vs. $p$.
As can be clearly seen, as $N$ increases, the
curves for E$(N_{1}/N)$ converge to the curves for the
results in the limit $N\to \infty$.
E$(N_{1}/N)$ for finite $N$ increases with $p$ up to a point and then decreases with $p$. As $N$ increases, the peak point for E$(N_{1}/N)$
moves rightward and converges to 1 in the limit $N\to\infty$.

To understand the mechanism for the bad performance
for large $p$, we derive the self-consistent equation
for $n_{1}$, which is the limiting value
of the expectation value of $N_{1}$, $n_{1}(t)\equiv \mbox{E}(N_{1}(t))$.
\[
n_{1}=\lim_{t\to\infty}n_{1}(t)=\lim_{t\to\infty} \sum_{n}P(n|t)n.
\]
It is given as
\begin{equation}
\frac{n_{1}}{N}=\left(1-\frac{q_{C}}{N}\right)\cdot
\frac{A+B-AP(0)}{q_{C}+(1-\frac{q_{C}}{N})(A+B)}\label{eq:N1_M12}.
\end{equation}
This is another expression for the result in Eq.(\ref{eq:N1_M1}).
In the limit $N\to \infty$, $P(0)=0$, and $q_{C}/N=0$.
We get the same result in Eq.~(\ref{eq:N1}).
The numerator in Eq.~(\ref{eq:N1_M12}) decreases as $P(0)$ increases.
As $p\to 1$ and $B=(1-p)q_{I}\to 0$, $P(0)$ increases to 1,
and the numerator reduces to 0.
As there is only one good arm in the system, agents are forced to
share the arm. If the arm changes to be bad, there is no agent
who knows a good arm, and copying fails.
If $p$ is large and $B$ is small,
agents cannot know a good arm for a long duration.
Thus, E$(N_{1}/N)$ becomes small compared with
the equilibrium value, where many good arms are known by agents and
copying always succeeds as $N_{1}>0$.

\section {\label{sec:MF}Mean Field Analysis for $N,M\to \infty$}
In this section, we study the distribution of $N$ agents in $M$ good arms
in the limit $N,M\to \infty$.
We denote the number of good arms with $k$ agents at $t$ as $M_{k}(t)$.
As there are $N$ agents, $k\in \{0,\cdots,N\}$.
In terms of $\vec{K}(t)$, $M_{k}(t)$ is written as
$M_{k}(t)=\sum_{m=1}^{M}\delta_{K_{m}(t),k}$.
$M_{k}(t)$ satisfies $\sum_{k=0}^{N}M_{k}(t)=M$.
In terms of $M_{k}(t)$, we can write $N_{1}(t)$ as $
N_{1}(t)=\sum_{k=1}^{N}M_{k}(t)\cdot k$.
The probabilistic rules for $M_{k}$ by an agent action are summarized as
\begin{eqnarray}
&&\mbox{Pr}(\Delta M_{0}=-1)=(1-x)B \cdot \frac{M_{0}}{M}, \nonumber \\
&&\mbox{Pr}(\Delta M_{0}=0)=1-\mbox{Pr}(\Delta M_{0}=-1), \nonumber \\
&&\mbox{Pr}(\Delta M_{k}=1)=(1-x)\left(A\cdot \frac{(k-1)M_{k-1}}{N_{1}}
+B\cdot \frac{M_{k-1}}{M}\right), \nonumber \\
&&\mbox{Pr}(\Delta M_{k}=-1)=
(1-x)\left(A\cdot \frac{k M_{k}}{N_{1}}
+B\frac{M_{k}}{M}\right), \nonumber \\
&&\mbox{Pr}(\Delta M_{k}=0)=1-(\mbox{Pr}(\Delta M_{k}=1)+\mbox{Pr}(\Delta M_{k}
=-1)), \nonumber \\
&&k\in \{1,2,\cdots,N-1\}
\nonumber \\
&&\mbox{Pr}(\Delta M_{N}=+1)=(1-x)\left(A\cdot \frac{(N-1)M_{N-1}}{N_{1}}
+B\cdot \frac{M_{N-1}}{M}\right),
\nonumber \\
&&\mbox{Pr}(\Delta M_{N}=0)=1-\mbox{Pr}(\Delta M_{N}=+1),
\nonumber \\
&&x\equiv N_{1}/N. \nonumber
\end{eqnarray}
In the rules, the terms with the coefficient $A$ correspond to
successful copying. They should be considered to be zero
if $N_{1}=0$.
$N_{1}=\sum_{k}M_{k}\cdot k$ obeys the next probabilistic rule.
\begin{eqnarray}
&&\mbox{Pr}(\Delta N_{1}=1)=(1-x)\left[A\cdot\left(\frac{N_{1}-N\cdot M_{N}}{N_{1}}\right)
+B\cdot \left(1-\frac{M_{N}}{M}\right)\right], \nonumber \\
&&\mbox{Pr}(\Delta N_{1}=0)=1-\mbox{Pr}(\Delta N_{1}=1).
\end{eqnarray}
By changing arms, $M_{k}$ changes with the binomial
probabilities
$\mbox{B}(M_{k},p_{c}/N)$:
\[
\mbox{Pr}(\Delta M_{k}=-\Delta m_{k})=
\binom{M_{k}}{\Delta m_{k}}\left(\frac{q_{C}}{N}\right)^{\Delta m_{k}}\left(1-\frac{q_{C}}{N}\right)^{M_{k}-\Delta m_{k}}.
\]
$M_{0}$ increases by $\sum_{k=1}^{N}\Delta m_{k}$, and
$N_{1}$ decreases by $\sum_{k=1}^{N}\Delta m_{k}\cdot k$.

We denote the expectation values of $M_{k}(t)$ and $N_{1}(t)$ as
$m_{k}(t)$ and $n_{1}(t)$, respectively.
The recursive relation for $m_{0}(t)$ is
\[
  m_{0}(t+1)=m_{0}(t)-\left(1-\frac{n_{1}}{N}\right)\cdot \left(1-\frac{q_{C}}{N}\right)\cdot B\cdot \frac{m_{0}(t)}{M}
  +\frac{q_{C}}{N}\cdot (M-m_{0}(t)).
  \]
  The second term on the right-hand side of
  the equation corresponds to a decrease by random search
  with the probability
  $(1-N_{1}/N)\cdot B\cdot M_{0}/M$. With the probability $q_{C}/N$,
  the arm changes to
  bad, and the net decrease is reduced by $(1-q_{C}/N)$.
  The third term corresponds to the change of good arms. There are
  $M-m_{0}$ good arms with nonzero agents, and they change to bad with the
  probability $q_{C}/N$.

The recursive relations for $m_{k},k\ge 1$ are
\begin{eqnarray}
  m_{k}(t+1)&=&\left(1-\frac{q_{C}}{N}\right)\left\{
  m_{k}(t)
  +\left(1-\frac{n_{1}(t)}{N}\right) \right.\nonumber \\
  &&\left.\left(A \cdot \frac{(k-1)m_{k-1}(t)-k m_{k}(t)}{n_{1}(t)}+
  B\cdot \frac{m_{k-1}(t)-m_{k}(t)}{M}\right)\right\}
  \,\,\, , \,\,\, k \in \{1,\cdots,N-1\},
  \nonumber \\
  m_{N}(t+1)&=&\left(1-\frac{q_{C}}{N}\right)\left\{m_{N}(t)+\left(1-\frac{n_{1}(t)}{N}\right)\left(A\cdot\frac{(N-1)m_{N-1}(t)}{n_{1}(t)}+B\cdot \frac{m_{N-1}(t)}{M}\right)
  \right\}. \nonumber
\end{eqnarray}
The prefactor $(1-q_{C}/N)$ on the right-hand side corresponds to the changes
of arms. In the bracket $\{,\}$, the terms proportional to $(1-n_{1}/N)$
correspond to a change by copying and a random search. As $q_{I}>0$ and $p<1$,
we can assume that $n_{1}(t)>0$ for $t>0$.
At $t=0$, $n_{1}(0)=0$, and the term proportional to $A$ should be zero.
 Using the above relations,
 we write the recursive relation for $n_{1}=\sum_{k}m_{k} k$ as
\begin{equation}
  n_{1}(t+1)=\left(1-\frac{q_{C}}{N}\right)\cdot \left\{
  n_{1}(t)+\left(1-\frac{n_{1}(t)}{N}\right)
  \left(A\cdot \frac{n_{1}(t)-N m_{N}(t)}{n_{1}(t)}+B\cdot
  \frac{M-m_{N}(t)}{M}\right)
\right\}. \label{eq:n1}
\end{equation}

\begin{figure}[htbp]
\begin{center}
\begin{tabular}{cc}
\includegraphics[width=8cm]{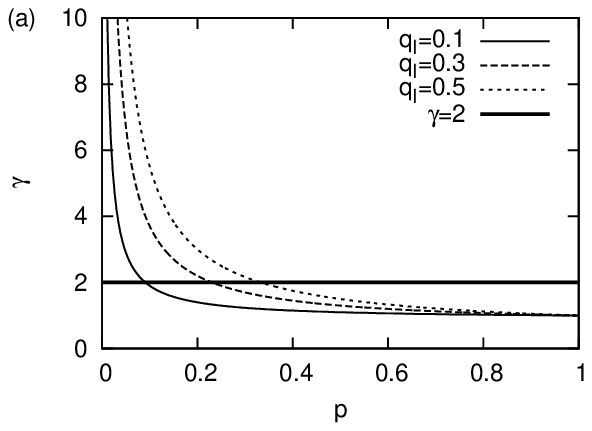}
&
\includegraphics[width=8cm]{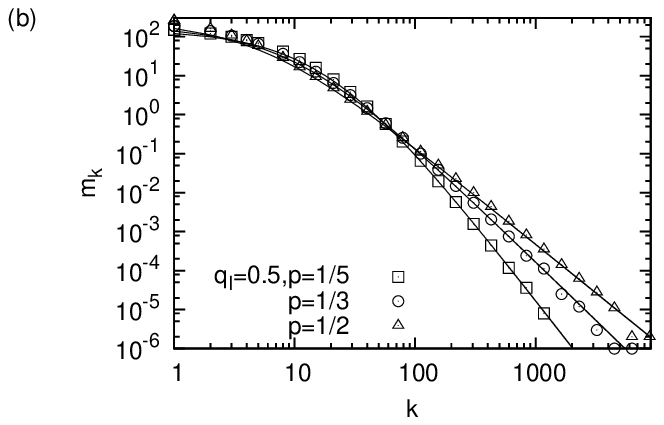}
\end{tabular}
\end{center}
\caption{\label{fig:gamma_Yule}
  (a)
    Plots of $p$ vs. $\gamma$ in Eq.~(\ref{eq:gamma}).
    $q_{c}=0.1,q_{O}=1$, and $q_{I}\in \{0.1,0.3,0.5\}$.
    (b) Plots of $m_{k}$ vs. $k$ of the numerical
    (symbols) 
    and theoretical results (solid curves)
    from Eq.~(\ref{eq:mk}).
    $q_{O}=1.0$ and $q_{I}=0.5$. $p=1/5,1/3,1/2$ and
    $\gamma=3,2,3/2$, respectively.
}
\end{figure}	

Here, we assume that the system is stationary for sufficiently large
$N$. The stationary value of $n_{1}(t)/N$, which we denote
$x$, is estimated by Eq.~(\ref{eq:N1}) as
\[
x=\lim_{t\to \infty}n_{1}(t)/N=\frac{A+B}{A+B+p_{c}}.
\]
$m_{k}(t)$ evolves to their stationary values, and
we write them as $m_{k}\equiv \lim_{t\to\infty}m_{k}(t)$.
We solve the above recursive relations by replacing
$m_{k}(t+1)$ and $m_{k}(t)$ with $m_{k}$.
$m_{0}$ is obtained as
\[
m_{0}=\frac{M}{1+\alpha/\gamma}\,\,\, ,\,\,\,
\alpha\equiv \frac{B\cdot N x}{A\cdot M}\,\,\, , \,\,\,
\gamma\equiv \frac{x q_{C}}{(1-x)(1-q_{C}/N)A}.
\]
$m_{k}$ and $m_{N}$ obey the following recursive relations:
\[
  m_{k}=\frac{(k-1)+\alpha}{k+\alpha+\gamma}\cdot m_{k-1} \,\,\, , \,\,\,
  m_{N}= \frac{(N-1)+\alpha}{\gamma}\cdot m_{N-1}.
\]
We obtain $m_{k}$ and $m_{N}$ as
\begin{eqnarray}
m_{k}&=&\frac{B(k+\alpha,\gamma+1)}{B(\alpha,\gamma+1)}\cdot m_{0}, \nonumber \\
m_{N}&=&\frac{B(N+\alpha,\gamma)}{B(\alpha,\gamma+1)}\cdot m_{0}
=\frac{N+\alpha+\gamma}{\gamma}\cdot\frac{B(N+\alpha,\gamma+1)}{B(\alpha,\gamma+1)}\cdot m_{0}. \label{eq:mk}
\end{eqnarray}
$m_{k},k<N$ obeys the Yule distribution~\cite{Yule:1925,Newman:2005}
(see Figure \ref{fig:gamma_Yule}). 

We take the limit $N,M\to \infty$ with $M/N=\beta$.
$\gamma$ is given as
\begin{equation}
\gamma=\frac{x}{1-x}\cdot \frac{q_{C}}{A}
=\frac{A+B}{A}=1+\frac{(1-p)q_{I}}{pq_{O}} \label{eq:gamma}.
\end{equation}
$\alpha$ is also written as
\[
\alpha=\frac{Bx}{A\beta}.
\]
We see that the power-law exponent for $m_{k}$
is given by $1+\gamma$.
\[
m_{k}\propto k^{-1-\gamma} =k^{-(2+\frac{(1-p)q_{I}}{pq_{O}})}.
\]
$\gamma$ decreases with $p$, as in Figure
\ref{fig:gamma_Yule}. 
As $\gamma$ is larger than one for $p<1$ and
and $m_{N}\propto (N+\alpha)^{-\gamma}$,
$m_{N}\to 0,Nm_{N}\to 0$ in the limit.
The recursive relation for $n_{1}$ in Eq.~(\ref{eq:n1})
then becomes an equilibrium
condition for $n_{1}$ when one replaces $n_{1}(t+1)$
 and $n_{1}(t)$ by $n_{1}$ as
\[
\frac{n_{1}}{N}=\left(1-\frac{p_{c}}{N}\right)\cdot \left\{
\frac{n_{1}}{N}+\left(1-\frac{n_{1}}{N}\right)\cdot \frac{A+B}{N}  \right\}.
\]
It reduces to the same equilibrium equation for $x=n_{1}/N$ in Eq.~(\ref{eq:Eq})
in the limit $N\to \infty$. The result is consistent with
the assumption that $n_{1}$ converges to the stationary value.

To estimate the sufficient system size $N_{s}$ for
the stationarity of the system,
we decompose the summation $\sum_{k}$ for
$x=n_{1}/N=\sum_{k=0}m_{k}k/N$ in the former part for $k\le N_{s}$
and the latter part for $k>N_{s}$ as
\[
x=\sum_{k=0}^{\infty}m_{k}k/N=\sum_{k=0}^{N_{s}}m_{k}k/N+\sum_{k=N_{s}+1}^{\infty}
m_{k}k/N.
\]
The latter part is then given as
\[
\sum_{k=N_{s}+1}^{\infty}m_{k}k/N=
\frac{B(\alpha+N_{s},\gamma+1)}{B(\alpha,\gamma+1)}
\frac{\alpha+\gamma+N_{s}}{\alpha+\gamma}
\frac{\alpha+N_{s}}{\gamma-1}\beta \propto N_{s}^{1-\gamma}.
\]
If we set $N_{s}=0$, as $x=\sum_{k=N_{s}+1}^{\infty}m_{k}k/N$,
we obtain the self-consistent equation
for $x$, which is the same as the equilibrium
condition for $n_{1}/N$ in Eq.~(\ref{eq:Eq}).
For the former term $\sum_{k=1}^{N_{s}}m_{k}k/N$ almost coincides with $x$, the latter term should be almost zero.
As $p\to 1$, $\gamma\to 1$, and the convergence of the latter term to zero
becomes remarkably slow.
We estimate the ratio of the latter term to $x$ as
\[
\frac{1}{x}\sum_{k=N_{s}+1}^{\infty}m_{k}k/N=
\frac{B(\alpha+N,\gamma+1)}{B(\alpha,\gamma+1)}
\cdot\frac{\alpha+\gamma+N}{\alpha+\gamma}\cdot
\frac{\alpha+N_{s}}{\alpha}\simeq
\frac{B(\alpha+N_{s},\gamma+1)}{B(\alpha,\gamma+1)}N_{s}^{2}.
\]
The ratio is proportional to $N_{s}^{1-\gamma}$, and it
converges to zero in the limit $N_{s}\to \infty$.
If the ratio is smaller than some small
value $\delta$, $N_{s}$ provides the adequate
system size up to the ratio $\delta$.
$N_{s}$ is then estimated as
\[
\frac{B(\alpha+N_{s},\gamma+1)}{B(\alpha,\gamma+1)}N_{s}^{2}
=\delta.
\]
Since the left-hand side behaves as $N_{s}^{1-\gamma}$,
we have $\ln N_{s} \propto \ln \delta/(\gamma-1)$.

We find that $x=n_{1}/N$ is given by the equilibrium value
in Eq.~(\ref{eq:N1}), and the agent distribution is scale-free, i.e.,
$m_{k}\sim (k+\alpha)^{-(\gamma+1)}$ if $N>N_{s}$.
As $p$ increases from zero to one, $x$ increases from
$q_{I}/(q_{I}+q_{C})$ to $q_{O}/(q_{O}+q_{C})$.
At the same time, $\gamma$ decreases to one, and
the adequate system size $N_{s}$ diverges.
If $N$ becomes smaller than $N_{s}$, the system is not in
equilibrium, and $N_{1}$ shows strong fluctuations.

\section{\label{sec:PTE}Phase Transition and Echo Chamber}
We estimate the fluctuation of $N_{1}$ around the stationary value
$n_{1}$ for $N>N_{s}$.
A good arm changes to bad with the probability $q_{C}/N$.
If the changed arm is known by $k$ agents, $\Delta N_{1}=-k$, and
$\Delta M_{k}=-1$. Here, we assume that $N$ is sufficiently large, and
we can neglect the change in $M_{k}$. We replace
$M_{k}$ with their
expectation values $m_{k}$ in Eq.~(\ref{eq:mk}).
Then, we write the probabilistic rule for $N_{1}\in [0,N]$ as
\begin{eqnarray}
&&\mbox{Pr}(\Delta N_{1}=1)=\left(1-\frac{N_{1}}{N}\right)(A+B), \nonumber \\
&&\mbox{Pr}(\Delta N_{1}=-k)=\frac{q_{C}}{N}\cdot m_{k}. \nonumber
\end{eqnarray}
The first equation corresponds to the action of a randomly chosen agent.
If $N_{1}=0$, we assume $\mbox{Pr}(\Delta N_{1}=1)=B$, as copying fails.
The second equation corresponds to the random change of arms.
Here, we assume that there occurs only one arm change in $M_{k}$
arms with $k$ agents. Since we assume $N$ is sufficiently large,
the assumption is reasonable.

The expectation value of $\Delta N_{1}$ is estimated as
\[
\mbox{E}(\Delta N_{1})=\left(1-\frac{N_{1}}{N}\right)(A+B)
-\frac{p_{c}}{N}\sum_{k=1}^{N_{1}}m_{k}\cdot k.
\]
As $N>N_{s}$, $\sum_{k=1}^{N}m_{k}k \simeq Nx$, and
$\mbox{E}(\Delta N_{1})$ vanishes for $N_{1}<N$, which
we denote as $N_{1,c}$.
If $\gamma>2$, $m_{k}$ decays rapidly, and the contribution
from the tail part in $\sum_{k=1}^{N}m_{k}k$ is negligible.
In this case, $N_{1,c}\simeq N x$, and $N_{1}$ fluctuates around $Nx$.
However, if $\gamma<2$, the contribution
from the tail part in $\sum_{k=1}^{N}m_{k}k$ is not negligible.
$N_{1,c}$ is larger than $Nx$.

\begin{figure}[htbp]
\begin{center}
\begin{tabular}{cc}
\includegraphics[width=8cm]{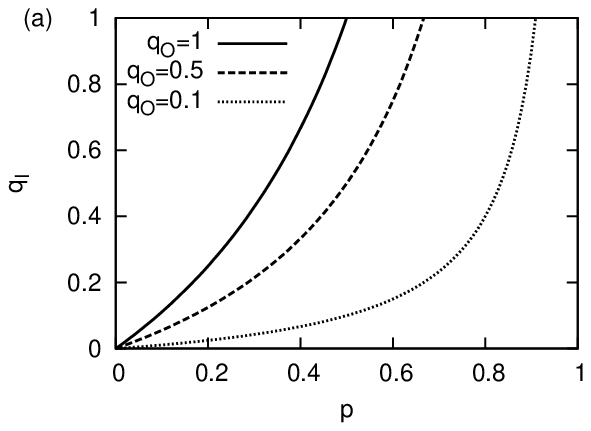}
&
\includegraphics[width=8cm]{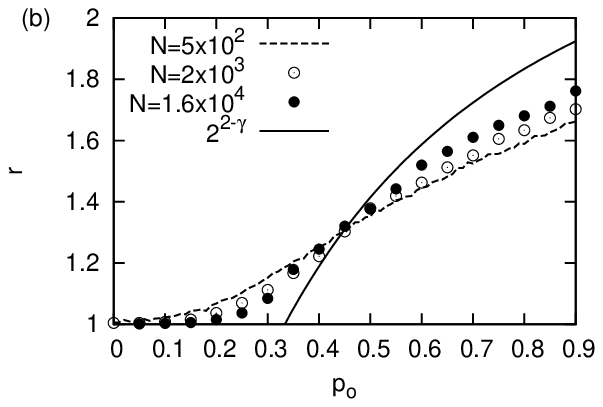}
\end{tabular}
\end{center}
\caption{\label{fig:phase}
  (a) Phase diagram in the $(p,q_{I})$ plane.
  Plots of $p_{c}$ vs. $q_{I}$ in Eq.~(\ref{eq:po_c}) for
  $q_{O}\in \{1,0.5,0.1\}$.
  (b) Plots of $r$ vs. $p$.
  $q_{C}=0.1,q_{I}=0.5,q_{O}=1,\beta=0.1$, and $N\in \{500,2000,16000\}$.
  The thin solid line corresponds to the limit
  behavior $2^{2-\gamma}$ for $p>p_{c}=1/3$.}
\end{figure}

We estimate the size of the fluctuation of $N_{1}$ around $N_{1,c}$.
We denote the variance of $N_{1}$ as $V_{N_{1}}(N)\equiv \mbox{Var}(N_{1})$,
where we write the system size dependence explicitly.
As there are $N$ agents, $V_{N_{1}}(N)$ is estimated by $N$ times
the variance of $\Delta N_{1}$. The variance of $N_{1}$ per
agent is then estimated as
\[
V_{N_{1}}(N)/N\equiv \mbox{E}((\Delta N_{1})^{2})
=(1-\frac{N_{1,c}}{N})(A+B)+
\sum_{k=1}^{N_{1,c}}m_{k} k^{2}.
\]
$m_{k}\sim k^{-1-\gamma}$, and the second term diverges as $N_{1,c}^{2-\gamma}$
for $\gamma\le 2$. As $N_{1,c}\propto N$, we have the scaling relation for
$V_{N_{1}}(N)$ as
\begin{equation}
V_{N_{1}}(N)/N
\propto
\begin{cases}
N^{2-\gamma}\,\,\,\, \mbox{for} \,\,\,\, \gamma < 2,   \\
\ln N  \,\,\,\, \mbox{for} \,\,\,\, \gamma=2, \\
N^{0}  \,\,\,\, \mbox{for} \,\,\,\, \gamma>2.
\end{cases}
\end{equation}
If $\gamma>2$, $V_{N_{1}}(N)/N$ does not depend on the system size $N$ and
is finite in the limit $N\to \infty$.
If $\gamma<2$,$V_{N_{1}}(N)/N$ diverges as
$N^{2-\gamma}$. For $\gamma=2$, we see logarithmic divergence in $V_{N_{1}}(N)$.
The system shows a phase transition as $\gamma$
passes between the two phases with a different asymptotic behavior
for $V_{N_{1}}(N)/N$. We denote the threshold value for $p$ as $p_{c}$,
which is given by
\begin{equation}
p_{c}=\frac{q_{I}}{q_{I}+q_{O}}.   \label{eq:po_c}
\end{equation}
If $p>p_{c}$, $\gamma<2$, and $V_{N_{1}}(N)/N \propto N^{2-\gamma}$.
If $p<p_{c}$, $\gamma>2$, and $V_{N_{1}}(N)/N\propto N^{0}$.
The left figure in Figure \ref{fig:phase} shows the phase boundary
in the $(p,q_{I})$ plane. We adopt $q_{O}\in \{1,0.5,0.1\}$.
In general, $p_{c}$ is an increasing function of $q_{I}$ and a decreasing
function of $q_{O}$. If $q_{I}=0.5$ and $q_{O}=1$, $p_{c}$ is estimated as
$1/3$.

To check the phase transition,
we study the ratio of $V_{N_{1}}(N)/N$ for $N=N$ and $N=2N$.
We denote the ratio as $r$.
\[
r\equiv \frac{V_{N_{1}}(2N)/2N}{V_{N_{1}}(N)/N}.
\]
If $\gamma >2$, $V_{N_{1}}(N)/N$ does not depends on $N$, and
$r=2^{0}=1$. If $\gamma<2$, $V_{N_{1}}(N)\propto N^{2-\gamma}$, and
$r=2^{2-\gamma}$.
Figure \ref{fig:phase}(b) plots
$r$ vs. $p$. We adopt $q_{I}=0.5$, and $q_{C}=0.1,q_{O}=1$ and $\beta=0.1$.
We have performed Monte Carlo simulations with $10^{6}$ MCS.
The system sizes $N$ are $N=5\times 10^{2},2\times 10^{3}$, and $1.6\times 10^{4}$.
If $p<p_{c}=1/3$, $r$ is almost one, and this suggests that
$V_{N_{1}}(N)/N$ does not depend on $N$. If $p>p_{c}$,
$r$ deviates from one, which means that $V_{N_{1}}(N)/N$ depends
on $N$. The numerical results for $r$ almost coincide with $2^{2-\gamma}$
if $p$ is not much larger than $p_{c}$ and $N=1.6\times 10^{4}$.
The scaling behavior for $V_{N_{1}}(N)$ holds for $N>N_{s}$.
In order to check the scaling behaviors for $p>>p_{c}$,
it is necessary to study a system with larger size $N$.

\begin{figure}[htbp]
\begin{center}
\begin{tabular}{cc}
\includegraphics[width=8cm]{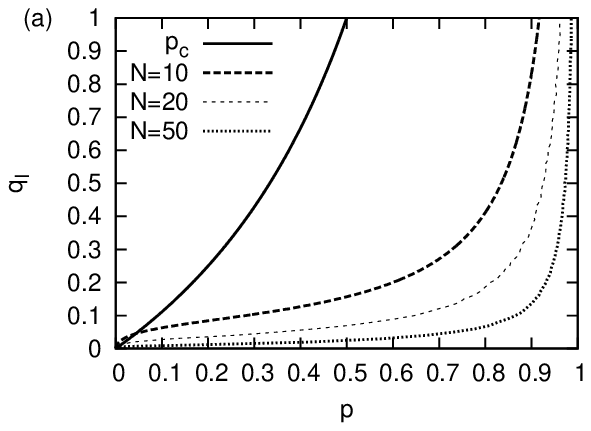}
&
\includegraphics[width=8cm]{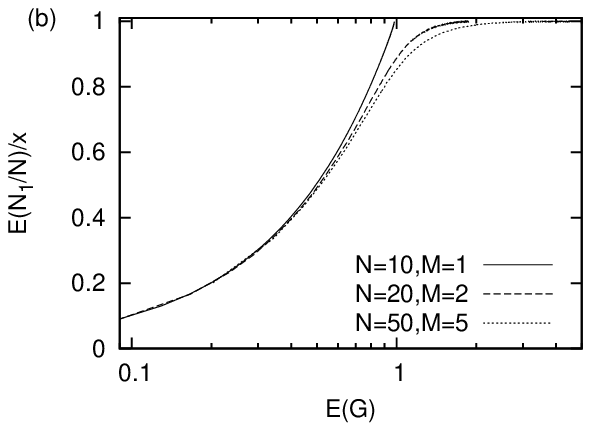}
\end{tabular}
\end{center}
\caption{\label{fig:echo}
  (a) Plot of $p_{c}$ and $p_{EC}$ vs. $q_{I}$ in the $(p,q_{I})$ plane.
  Plots of $p_{c}$ vs. $q_{I}$ in Eq.~(\ref{eq:po_c}) for
  $q_{O}=1$ in thick solid line. 
  $p_{EC}$ is defined as the threshold value where
  $\mbox{E}(N_{1}/N)/x=0.9$.
  We adopt $q_{c}=0.1,q_{I}=0.5,\beta=0.1$ and $N\in \{10,20,50\}$.
  (b) Plot of $\mbox{E}(N_{1}/N)/x$ vs.
  $\mbox{E}(G)\equiv \mbox{E}\left(\sum_{m}(1-\delta_{K_{m},0})\right)$.
  We adopt the same parameters with the left figure.
}
\end{figure}	

At last, we study the ``Echo chamber'' state of the system.
For large $p$, $\mbox{E}(N_{1}/N)$ becomes a decreasing function of $p$.
We study how the performance of the agents is deteriorated for large $p$.
We define the Echo chamber state where the ratio
$\mbox{E}(N_{1}/N)/x$ is smaller than $0.9$.
We denote the threshold value as $p_{EC}$.
We numerically estimate $p_{EC}$ and show the results in Figure
\ref{fig:echo}a.
We set $p_{c}=0.1,\beta=0.1$, and $N\in \{10,20,50\}$.
We see that the Echo chamber state is in the phase $p>p_{c}$,
except for $N=10,M=1$, and $q_{I}\simeq 0$.

In Figure \ref{fig:echo}b, we plot $\mbox{E}(N_{1}/N)/x$ vs.
the average number of good arms 
known by agents, which is denoted as $\mbox{E}(G)$.
$\mbox{E}(G)$ is estimated as
\[
\mbox{E}(G)\equiv \mbox{E}\left(\sum_{m}(1-\delta_{K_{m},0})\right).
\]
In the ``Echo chamber'' and $p>p_{EC}>p_{c}$, $\mbox{E}(N_{1}/N)/x$ is less than 0.9.
In this case, as Figure \ref{fig:echo}b shows,
the average number of known good arms is
less than one. This suggests that only one good arm is shared by
all agents on average in the ``Echo chamber'' state.

\section{\label{sec:Con} Conclusions}
We study a simple model for social learning agents in
an rMAB. There are $N$ agents and
the bandit has $M$ good arms that change to bad with the probability $q_{c}/N$.
If the agents do not know a good arm, they look for it by a
random search (with the success probability $q_{I}$) or
copy the information of other agents' good arms (with the success probability
$q_{O}$) with the probabilities $1-p$ or $p$, respectively.
We summarize the results below.

\begin{enumerate}
\item $N\to \infty$ limit: The system is in equilibrium, and
  the number of agents who know a good arm $N_{1}$ is balanced
  with the number of the agents who do not $N_{0}$.
The ratio $N_{1}/N$ becomes an increasing function of
$p$ for $p<1$ and $q_{I}<q_{O}$.

\item $M=1$ case: If $N$ is finite, $\mbox{E}(N_{1})/N$ decreases with $p$
for large $p$. There is only one good arm, and
agents cannot find a good one after the shared good arm changes
to be bad if $p$ is large.

\item $N,M\to \infty$ limit:
  The stationary distribution of agents in
  good arms $\{m_{k}\}$ is described by the Yule distribution
$m_{k}\sim k^{-1-\gamma}$, and  $\gamma=1+\frac{(1-p)q_{I}}{pq_{O}}$.
$N_{1}/N$ fluctuates around the equilibrium value in Eq.~(\ref{eq:N1}).
The system shows a phase transition at $p_{c}=q_{I}/(q_{I}+q_{O})$,
 and the asymptotic behavior of
 $\mbox{Var}(N_{1})/N$ changes.
 If $p<p_{c}$, $\mbox{Var}(N_{1})/N$ is finite in the limit $N\to \infty$.
 If $p>p_{c}$, $\mbox{Var}(N_{1})/N \propto N^{2-\gamma}$ and diverges
 with $N$.

\item $N,M<\infty$ case: There is a threshold value $N_{s}$
  about the system size $N$, and
  $N_{s}$ behaves as $\ln N_{s}\propto 1/(1-\gamma)$.
  If $N>N_{s}$, the system is in equilibrium and multiple good
  arms are known by agents. $\mbox{E}(N_{1})/N$ is almost the
  same as the equilibrium value.
  If $p<p_{c}$, $N_{s}$ is small, and $N>N_{s}$ usually holds.
  If $p>p_{c}$, $N_{s}$ becomes very large.
  If $N<N_{s}$, all agents tend to share one good arm, as in the case of $M=1$, and $\mbox{E}(N_{1})/N$ decreases with $p$, which is
 the ``Echo chamber'' state of social learning collectives.
\end{enumerate}

At last, we point out several future problems.
The first problem is the empirical verification of the
predictions of the model. The model predicts that the social learning agents
show a phase transition, and
the agents’ distribution has a scale-free nature.
In general, the performance of agents improves with $p$, and
the power-law exponent decreases with $p$.
These predictions should be studied
with the tournament results \cite{Rendell:2010}.
Experimental study is also an interesting research direction.
Thus far, there have been some experimental studies
of human collectives in an MAB \cite{Toyokawa:2014} and
rMAB\cite{Kameda:2002,Yoshida:2016}.
In these studies, the collective intelligence by social learning
has been studied. This paper provides a new experimental design
from the viewpoint of a phase transition.
The second problem is the network effect. In this paper, we treat
the case where the agents can copy any agent's arm. In reality,
agents should have limited access to other agents, and
the accessibility should be incorporated as a network\cite{Rendell:2009}.
The information of an arm is transmitted over the network, and
there is a possibility of a phase with multiple arms, even
for large $p$.
As for the third problem, a multi-armed bandit with arms having different payoffs should be studied. In the setting of the agent tournament, the payoff
obeys an exponential distribution, and an arm with high return is rare
\cite{Rendell:2010}.
As it is natural that an agent exploits the
arm or searches for a new arm depending on the return of his arm,
the model should incorporate the
exit of agents from good arms. In an evolving network model, incorporation of
the fitness of nodes leads to the Bose condensation of edges
\cite{Bianconi:2001-2}.
The collective behaviors of agents in the rMAB with
arms having different payoffs might be an interesting problem.

\begin{acknowledgments}
  One of the authors (S.M.) thanks Shunsuke Yoshida
  for useful discussions.
\end{acknowledgments}

\bibliography{myref201610}

\providecommand{\noopsort}[1]{}\providecommand{\singleletter}[1]{#1}%
\begin{thebibliography}{41}%
\makeatletter
\providecommand \@ifxundefined [1]{%
 \@ifx{#1\undefined}
}%
\providecommand \@ifnum [1]{%
 \ifnum #1\expandafter \@firstoftwo
 \else \expandafter \@secondoftwo
 \fi
}%
\providecommand \@ifx [1]{%
 \ifx #1\expandafter \@firstoftwo
 \else \expandafter \@secondoftwo
 \fi
}%
\providecommand \natexlab [1]{#1}%
\providecommand \enquote  [1]{``#1''}%
\providecommand \bibnamefont  [1]{#1}%
\providecommand \bibfnamefont [1]{#1}%
\providecommand \citenamefont [1]{#1}%
\providecommand \href@noop [0]{\@secondoftwo}%
\providecommand \href [0]{\begingroup \@sanitize@url \@href}%
\providecommand \@href[1]{\@@startlink{#1}\@@href}%
\providecommand \@@href[1]{\endgroup#1\@@endlink}%
\providecommand \@sanitize@url [0]{\catcode `\\12\catcode `\$12\catcode
  `\&12\catcode `\#12\catcode `\^12\catcode `\_12\catcode `\%12\relax}%
\providecommand \@@startlink[1]{}%
\providecommand \@@endlink[0]{}%
\providecommand \url  [0]{\begingroup\@sanitize@url \@url }%
\providecommand \@url [1]{\endgroup\@href {#1}{\urlprefix }}%
\providecommand \urlprefix  [0]{URL }%
\providecommand \Eprint [0]{\href }%
\providecommand \doibase [0]{http://dx.doi.org/}%
\providecommand \selectlanguage [0]{\@gobble}%
\providecommand \bibinfo  [0]{\@secondoftwo}%
\providecommand \bibfield  [0]{\@secondoftwo}%
\providecommand \translation [1]{[#1]}%
\providecommand \BibitemOpen [0]{}%
\providecommand \bibitemStop [0]{}%
\providecommand \bibitemNoStop [0]{.\EOS\space}%
\providecommand \EOS [0]{\spacefactor3000\relax}%
\providecommand \BibitemShut  [1]{\csname bibitem#1\endcsname}%
\let\auto@bib@innerbib\@empty
\bibitem [{\citenamefont {Pentland}(2014)}]{Pentland:2014}%
  \BibitemOpen
  \bibfield  {author} {\bibinfo {author} {\bibfnamefont {A.}~\bibnamefont
  {Pentland}},\ }\href@noop {} {\emph {\bibinfo {title} {Social Physics: How
  good ideas spread}}}\ (\bibinfo  {publisher} {Penguin Press},\ \bibinfo
  {year} {2014})\BibitemShut {NoStop}%
\bibitem [{\citenamefont {Ormerod}(2012)}]{Ormerod:2012}%
  \BibitemOpen
  \bibfield  {author} {\bibinfo {author} {\bibfnamefont {P.}~\bibnamefont
  {Ormerod}},\ }\href@noop {} {\emph {\bibinfo {title} {Positive Linking}}}\
  (\bibinfo  {publisher} {Faber \& Faber},\ \bibinfo {year} {2012})\BibitemShut
  {NoStop}%
\bibitem [{\citenamefont {Mantegna}\ and\ \citenamefont
  {Stanley}(2007)}]{Mantegna:2008}%
  \BibitemOpen
  \bibfield  {author} {\bibinfo {author} {\bibfnamefont {R.~N.}\ \bibnamefont
  {Mantegna}}\ and\ \bibinfo {author} {\bibfnamefont {H.~E.}\ \bibnamefont
  {Stanley}},\ }\href@noop {} {\emph {\bibinfo {title} {Introduction to
  Econophysics: Correlations and Complexity in Finance}}}\ (\bibinfo
  {publisher} {Cambridge University Press, Cambridge},\ \bibinfo {year}
  {2007})\BibitemShut {NoStop}%
\bibitem [{\citenamefont {Cont}\ and\ \citenamefont
  {Bouchaud}(2000)}]{Cont:2000}%
  \BibitemOpen
  \bibfield  {author} {\bibinfo {author} {\bibfnamefont {R.}~\bibnamefont
  {Cont}}\ and\ \bibinfo {author} {\bibfnamefont {J.}~\bibnamefont
  {Bouchaud}},\ }\href@noop {} {\bibfield  {journal} {\bibinfo  {journal}
  {Macroecon. Dynam.}\ }\textbf {\bibinfo {volume} {4}},\ \bibinfo {pages}
  {170} (\bibinfo {year} {2000})}\BibitemShut {NoStop}%
\bibitem [{\citenamefont {Lux}(1995)}]{Lux:1995}%
  \BibitemOpen
  \bibfield  {author} {\bibinfo {author} {\bibfnamefont {T.}~\bibnamefont
  {Lux}},\ }\href@noop {} {\bibfield  {journal} {\bibinfo  {journal} {Econ.
  J.}\ }\textbf {\bibinfo {volume} {105}},\ \bibinfo {pages} {881} (\bibinfo
  {year} {1995})}\BibitemShut {NoStop}%
\bibitem [{\citenamefont {Fernandez-Gracia}\ \emph {et~al.}(2014)\citenamefont
  {Fernandez-Gracia}, \citenamefont {Suchecki}, \citenamefont {Ramasco},
  \citenamefont {Miguel},\ and\ \citenamefont {Egu\'{i}luz}}]{Gracia:2014}%
  \BibitemOpen
  \bibfield  {author} {\bibinfo {author} {\bibfnamefont {J.}~\bibnamefont
  {Fernandez-Gracia}}, \bibinfo {author} {\bibfnamefont {K.}~\bibnamefont
  {Suchecki}}, \bibinfo {author} {\bibfnamefont {J.~J.}\ \bibnamefont
  {Ramasco}}, \bibinfo {author} {\bibfnamefont {M.~S.}\ \bibnamefont {Miguel}},
  \ and\ \bibinfo {author} {\bibfnamefont {V.~M.}\ \bibnamefont
  {Egu\'{i}luz}},\ }\href@noop {} {\bibfield  {journal} {\bibinfo  {journal}
  {Phys.Rev.Lett.}\ }\textbf {\bibinfo {volume} {112}},\ \bibinfo {pages}
  {158701} (\bibinfo {year} {2014})}\BibitemShut {NoStop}%
\bibitem [{\citenamefont {K.N.Laland}(2004)}]{Laland:2004}%
  \BibitemOpen
  \bibfield  {author} {\bibinfo {author} {\bibnamefont {K.N.Laland}},\
  }\href@noop {} {\bibfield  {journal} {\bibinfo  {journal} {Learn. Behav.}\
  }\textbf {\bibinfo {volume} {32}},\ \bibinfo {pages} {4} (\bibinfo {year}
  {2004})}\BibitemShut {NoStop}%
\bibitem [{\citenamefont {J.Kendal}\ \emph {et~al.}(2009)\citenamefont
  {J.Kendal}, \citenamefont {L.-A.Giraldeau},\ and\ \citenamefont
  {K.N.Laland}}]{Kendal:2009}%
  \BibitemOpen
  \bibfield  {author} {\bibinfo {author} {\bibnamefont {J.Kendal}}, \bibinfo
  {author} {\bibnamefont {L.-A.Giraldeau}}, \ and\ \bibinfo {author}
  {\bibnamefont {K.N.Laland}},\ }\href@noop {} {\bibfield  {journal} {\bibinfo
  {journal} {J.Theor.Biol.}\ }\textbf {\bibinfo {volume} {260}},\ \bibinfo
  {pages} {210} (\bibinfo {year} {2009})}\BibitemShut {NoStop}%
\bibitem [{\citenamefont {Kirman}(2010)}]{Kirman:2010}%
  \BibitemOpen
  \bibfield  {author} {\bibinfo {author} {\bibfnamefont {A.}~\bibnamefont
  {Kirman}},\ }\href@noop {} {\emph {\bibinfo {title} {Complex Economics:
  Individual and Collective Rationality}}}\ (\bibinfo  {publisher}
  {Routledge},\ \bibinfo {year} {2010})\BibitemShut {NoStop}%
\bibitem [{\citenamefont {D.Kahneman}(2003)}]{Kahneman:2003}%
  \BibitemOpen
  \bibfield  {author} {\bibinfo {author} {\bibnamefont {D.Kahneman}},\
  }\href@noop {} {\bibfield  {journal} {\bibinfo  {journal} {Am. Econ. Rev.}\
  }\textbf {\bibinfo {volume} {93}},\ \bibinfo {pages} {1449} (\bibinfo {year}
  {2003})}\BibitemShut {NoStop}%
\bibitem [{\citenamefont {Rendell}\ \emph {et~al.}(2011)\citenamefont
  {Rendell}, \citenamefont {Fogarty}, \citenamefont {Hoppitt}, \citenamefont
  {Morgan}, \citenamefont {Webster},\ and\ \citenamefont
  {Laland}}]{Rendell:2011}%
  \BibitemOpen
  \bibfield  {author} {\bibinfo {author} {\bibfnamefont {L.}~\bibnamefont
  {Rendell}}, \bibinfo {author} {\bibfnamefont {L.}~\bibnamefont {Fogarty}},
  \bibinfo {author} {\bibfnamefont {W.}~\bibnamefont {Hoppitt}}, \bibinfo
  {author} {\bibfnamefont {T.}~\bibnamefont {Morgan}}, \bibinfo {author}
  {\bibfnamefont {M.}~\bibnamefont {Webster}}, \ and\ \bibinfo {author}
  {\bibfnamefont {K.}~\bibnamefont {Laland}},\ }\href@noop {} {\bibfield
  {journal} {\bibinfo  {journal} {Trends Cogn. Sci.}\ }\textbf {\bibinfo
  {volume} {15}},\ \bibinfo {pages} {68} (\bibinfo {year} {2011})}\BibitemShut
  {NoStop}%
\bibitem [{\citenamefont {Rendell}\ \emph
  {et~al.}(2010{\natexlab{a}})\citenamefont {Rendell}, \citenamefont {Boyd},
  \citenamefont {Cownden}, \citenamefont {Enquist}, \citenamefont {Eriksson},
  \citenamefont {Feldman}, \citenamefont {Fogarty}, \citenamefont {Ghirlanda},
  \citenamefont {Lillicrap},\ and\ \citenamefont {Laland}}]{Rendell:2010}%
  \BibitemOpen
  \bibfield  {author} {\bibinfo {author} {\bibfnamefont {L.}~\bibnamefont
  {Rendell}}, \bibinfo {author} {\bibfnamefont {R.}~\bibnamefont {Boyd}},
  \bibinfo {author} {\bibfnamefont {D.}~\bibnamefont {Cownden}}, \bibinfo
  {author} {\bibfnamefont {M.}~\bibnamefont {Enquist}}, \bibinfo {author}
  {\bibfnamefont {K.}~\bibnamefont {Eriksson}}, \bibinfo {author}
  {\bibfnamefont {M.~W.}\ \bibnamefont {Feldman}}, \bibinfo {author}
  {\bibfnamefont {L.}~\bibnamefont {Fogarty}}, \bibinfo {author} {\bibfnamefont
  {S.}~\bibnamefont {Ghirlanda}}, \bibinfo {author} {\bibfnamefont
  {T.}~\bibnamefont {Lillicrap}}, \ and\ \bibinfo {author} {\bibfnamefont
  {K.~N.}\ \bibnamefont {Laland}},\ }\href@noop {} {\bibfield  {journal}
  {\bibinfo  {journal} {Science}\ }\textbf {\bibinfo {volume} {328}},\ \bibinfo
  {pages} {208} (\bibinfo {year} {2010}{\natexlab{a}})}\BibitemShut {NoStop}%
\bibitem [{\citenamefont {Galam}(2008)}]{Galam:2008}%
  \BibitemOpen
  \bibfield  {author} {\bibinfo {author} {\bibfnamefont {S.}~\bibnamefont
  {Galam}},\ }\href@noop {} {\bibfield  {journal} {\bibinfo  {journal} {Int. J.
  Mod. Phys. C}\ }\textbf {\bibinfo {volume} {19}},\ \bibinfo {pages} {409}
  (\bibinfo {year} {2008})}\BibitemShut {NoStop}%
\bibitem [{\citenamefont {Castellano}\ \emph {et~al.}(2009)\citenamefont
  {Castellano}, \citenamefont {Fortunato},\ and\ \citenamefont
  {Loreto}}]{Castellano:2009}%
  \BibitemOpen
  \bibfield  {author} {\bibinfo {author} {\bibfnamefont {C.}~\bibnamefont
  {Castellano}}, \bibinfo {author} {\bibfnamefont {S.}~\bibnamefont
  {Fortunato}}, \ and\ \bibinfo {author} {\bibfnamefont {V.}~\bibnamefont
  {Loreto}},\ }\href@noop {} {\bibfield  {journal} {\bibinfo  {journal}
  {Rev.Mod.Phys.}\ }\textbf {\bibinfo {volume} {81}},\ \bibinfo {pages} {591}
  (\bibinfo {year} {2009})}\BibitemShut {NoStop}%
\bibitem [{\citenamefont {Hisakado}\ and\ \citenamefont
  {Mori}(2011)}]{Hisakado:2011}%
  \BibitemOpen
  \bibfield  {author} {\bibinfo {author} {\bibfnamefont {M.}~\bibnamefont
  {Hisakado}}\ and\ \bibinfo {author} {\bibfnamefont {S.}~\bibnamefont
  {Mori}},\ }\href@noop {} {\bibfield  {journal} {\bibinfo  {journal} {J. Phys.
  A}\ }\textbf {\bibinfo {volume} {44}},\ \bibinfo {pages} {275204} (\bibinfo
  {year} {2011})}\BibitemShut {NoStop}%
\bibitem [{\citenamefont {Mori}\ and\ \citenamefont
  {Hisakado}(2015)}]{Mori:2015-2}%
  \BibitemOpen
  \bibfield  {author} {\bibinfo {author} {\bibfnamefont {S.}~\bibnamefont
  {Mori}}\ and\ \bibinfo {author} {\bibfnamefont {M.}~\bibnamefont
  {Hisakado}},\ }\href@noop {} {\bibfield  {journal} {\bibinfo  {journal}
  {Phys.Rev. E}\ }\textbf {\bibinfo {volume} {92}},\ \bibinfo {pages} {052112}
  (\bibinfo {year} {2015})}\BibitemShut {NoStop}%
\bibitem [{\citenamefont {Bikhchandani}\ \emph {et~al.}(1992)\citenamefont
  {Bikhchandani}, \citenamefont {Hirshleifer},\ and\ \citenamefont
  {Welch}}]{Bikchandani:1992}%
  \BibitemOpen
  \bibfield  {author} {\bibinfo {author} {\bibfnamefont {S.}~\bibnamefont
  {Bikhchandani}}, \bibinfo {author} {\bibfnamefont {D.}~\bibnamefont
  {Hirshleifer}}, \ and\ \bibinfo {author} {\bibfnamefont {I.}~\bibnamefont
  {Welch}},\ }\href@noop {} {\bibfield  {journal} {\bibinfo  {journal} {J.
  Polit. Econ.}\ }\textbf {\bibinfo {volume} {100}},\ \bibinfo {pages} {992}
  (\bibinfo {year} {1992})}\BibitemShut {NoStop}%
\bibitem [{\citenamefont {Devenow}\ and\ \citenamefont
  {Welch}(1996)}]{Devenow:1996}%
  \BibitemOpen
  \bibfield  {author} {\bibinfo {author} {\bibfnamefont {A.}~\bibnamefont
  {Devenow}}\ and\ \bibinfo {author} {\bibfnamefont {I.}~\bibnamefont
  {Welch}},\ }\href@noop {} {\bibfield  {journal} {\bibinfo  {journal} {Euro.
  Econ. Rev.}\ }\textbf {\bibinfo {volume} {40}},\ \bibinfo {pages} {603}
  (\bibinfo {year} {1996})}\BibitemShut {NoStop}%
\bibitem [{\citenamefont {Mori}\ \emph {et~al.}(2012)\citenamefont {Mori},
  \citenamefont {Hisakado},\ and\ \citenamefont {Takahashi}}]{Mori:2012}%
  \BibitemOpen
  \bibfield  {author} {\bibinfo {author} {\bibfnamefont {S.}~\bibnamefont
  {Mori}}, \bibinfo {author} {\bibfnamefont {M.}~\bibnamefont {Hisakado}}, \
  and\ \bibinfo {author} {\bibfnamefont {T.}~\bibnamefont {Takahashi}},\
  }\href@noop {} {\bibfield  {journal} {\bibinfo  {journal} {Phys. Rev. E}\
  }\textbf {\bibinfo {volume} {86}},\ \bibinfo {pages} {026109} (\bibinfo
  {year} {2012})}\BibitemShut {NoStop}%
\bibitem [{\citenamefont {Hino}\ \emph {et~al.}(2016)\citenamefont {Hino},
  \citenamefont {Irie}, \citenamefont {Hisakado}, \citenamefont {Takahashi},\
  and\ \citenamefont {S.Mori}}]{Hino:2016}%
  \BibitemOpen
  \bibfield  {author} {\bibinfo {author} {\bibfnamefont {M.}~\bibnamefont
  {Hino}}, \bibinfo {author} {\bibfnamefont {Y.}~\bibnamefont {Irie}}, \bibinfo
  {author} {\bibfnamefont {M.}~\bibnamefont {Hisakado}}, \bibinfo {author}
  {\bibfnamefont {T.}~\bibnamefont {Takahashi}}, \ and\ \bibinfo {author}
  {\bibnamefont {S.Mori}},\ }\href@noop {} {\bibfield  {journal} {\bibinfo
  {journal} {J.Phys.Soc.Jpn.}\ }\textbf {\bibinfo {volume} {85}},\ \bibinfo
  {pages} {034002} (\bibinfo {year} {2016})}\BibitemShut {NoStop}%
\bibitem [{\citenamefont {Austen-Smith}\ and\ \citenamefont
  {Banks}(1996)}]{Austen-Smith:1996}%
  \BibitemOpen
  \bibfield  {author} {\bibinfo {author} {\bibfnamefont {D.}~\bibnamefont
  {Austen-Smith}}\ and\ \bibinfo {author} {\bibfnamefont {J.~S.}\ \bibnamefont
  {Banks}},\ }\href@noop {} {\bibfield  {journal} {\bibinfo  {journal} {Am.
  Pol. Sci. Rev.}\ }\textbf {\bibinfo {volume} {90}},\ \bibinfo {pages} {34}
  (\bibinfo {year} {1996})}\BibitemShut {NoStop}%
\bibitem [{\citenamefont {Surowiecki}(2004)}]{Surowiecki:2004}%
  \BibitemOpen
  \bibfield  {author} {\bibinfo {author} {\bibfnamefont {J.}~\bibnamefont
  {Surowiecki}},\ }\href@noop {} {\emph {\bibinfo {title} {The Wisdom of
  Crowds}}}\ (\bibinfo  {publisher} {Doubleday, New York},\ \bibinfo {year}
  {2004})\BibitemShut {NoStop}%
\bibitem [{\citenamefont {Conradt}\ and\ \citenamefont
  {Lisst}(2009)}]{Conradt:2009}%
  \BibitemOpen
  \bibfield  {author} {\bibinfo {author} {\bibfnamefont {L.}~\bibnamefont
  {Conradt}}\ and\ \bibinfo {author} {\bibfnamefont {C.}~\bibnamefont
  {Lisst}},\ }\href@noop {} {\bibfield  {journal} {\bibinfo  {journal} {Phil.
  Trans. R. Soc.}\ }\textbf {\bibinfo {volume} {B364}},\ \bibinfo {pages} {719}
  (\bibinfo {year} {2009})}\BibitemShut {NoStop}%
\bibitem [{\citenamefont {Sumpter}\ and\ \citenamefont
  {Pratt}(2009)}]{Sumpter:2009}%
  \BibitemOpen
  \bibfield  {author} {\bibinfo {author} {\bibfnamefont {D.}~\bibnamefont
  {Sumpter}}\ and\ \bibinfo {author} {\bibfnamefont {S.~C.}\ \bibnamefont
  {Pratt}},\ }\href@noop {} {\bibfield  {journal} {\bibinfo  {journal} {Phil.
  Trans. R. Soc.}\ }\textbf {\bibinfo {volume} {B364}},\ \bibinfo {pages} {743}
  (\bibinfo {year} {2009})}\BibitemShut {NoStop}%
\bibitem [{\citenamefont {L.-A.Giraldeau}\ \emph {et~al.}(2002)\citenamefont
  {L.-A.Giraldeau}, \citenamefont {T.J.Valone},\ and\ \citenamefont
  {J.J.Templeton}}]{Giraldeau:2002}%
  \BibitemOpen
  \bibfield  {author} {\bibinfo {author} {\bibnamefont {L.-A.Giraldeau}},
  \bibinfo {author} {\bibnamefont {T.J.Valone}}, \ and\ \bibinfo {author}
  {\bibnamefont {J.J.Templeton}},\ }\href@noop {} {\bibfield  {journal}
  {\bibinfo  {journal} {Philos.Trans.R.Soc.London Ser.B}\ }\textbf {\bibinfo
  {volume} {357}},\ \bibinfo {pages} {1559} (\bibinfo {year}
  {2002})}\BibitemShut {NoStop}%
\bibitem [{\citenamefont {Rogers}(1988)}]{Rogers:1988}%
  \BibitemOpen
  \bibfield  {author} {\bibinfo {author} {\bibfnamefont {A.}~\bibnamefont
  {Rogers}},\ }\href@noop {} {\bibfield  {journal} {\bibinfo  {journal} {Am.
  Anthroplo.}\ }\textbf {\bibinfo {volume} {90}},\ \bibinfo {pages} {819}
  (\bibinfo {year} {1988})}\BibitemShut {NoStop}%
\bibitem [{\citenamefont {Enquist}\ \emph {et~al.}(2007)\citenamefont
  {Enquist}, \citenamefont {Eriksson},\ and\ \citenamefont
  {Ghirlanda}}]{Enquist:2007}%
  \BibitemOpen
  \bibfield  {author} {\bibinfo {author} {\bibfnamefont {M.}~\bibnamefont
  {Enquist}}, \bibinfo {author} {\bibfnamefont {K.}~\bibnamefont {Eriksson}}, \
  and\ \bibinfo {author} {\bibfnamefont {G.}~\bibnamefont {Ghirlanda}},\
  }\href@noop {} {\bibfield  {journal} {\bibinfo  {journal} {Am. Anthroplo.}\
  }\textbf {\bibinfo {volume} {109}},\ \bibinfo {pages} {727} (\bibinfo {year}
  {2007})}\BibitemShut {NoStop}%
\bibitem [{\citenamefont {Pemantle}(2007)}]{Pemantle:2007}%
  \BibitemOpen
  \bibfield  {author} {\bibinfo {author} {\bibfnamefont {R.}~\bibnamefont
  {Pemantle}},\ }\href@noop {} {\bibfield  {journal} {\bibinfo  {journal}
  {Pobab. Surv.}\ }\textbf {\bibinfo {volume} {4}},\ \bibinfo {pages} {1}
  (\bibinfo {year} {2007})}\BibitemShut {NoStop}%
\bibitem [{\citenamefont {Barab\'{a}si}\ and\ \citenamefont
  {Albert}(1999)}]{Barabasi:1999}%
  \BibitemOpen
  \bibfield  {author} {\bibinfo {author} {\bibfnamefont {A.}~\bibnamefont
  {Barab\'{a}si}}\ and\ \bibinfo {author} {\bibfnamefont {R.}~\bibnamefont
  {Albert}},\ }\href@noop {} {\bibfield  {journal} {\bibinfo  {journal}
  {Science}\ }\textbf {\bibinfo {volume} {286}},\ \bibinfo {pages} {509}
  (\bibinfo {year} {1999})}\BibitemShut {NoStop}%
\bibitem [{\citenamefont {S.N.Dorogovtsev}\ and\ \citenamefont
  {J.F.F.Mendes}(2000)}]{Dorogovtsev:2000}%
  \BibitemOpen
  \bibfield  {author} {\bibinfo {author} {\bibnamefont {S.N.Dorogovtsev}}\ and\
  \bibinfo {author} {\bibnamefont {J.F.F.Mendes}},\ }\href@noop {} {\bibfield
  {journal} {\bibinfo  {journal} {Phys.Rev.E}\ }\textbf {\bibinfo {volume}
  {62}},\ \bibinfo {pages} {1842} (\bibinfo {year} {2000})}\BibitemShut
  {NoStop}%
\bibitem [{\citenamefont {F.D.J.Neiman}(1995)}]{Neiman:1995}%
  \BibitemOpen
  \bibfield  {author} {\bibinfo {author} {\bibnamefont {F.D.J.Neiman}},\
  }\href@noop {} {\bibfield  {journal} {\bibinfo  {journal} {Am. Aniquity}\
  }\textbf {\bibinfo {volume} {60}},\ \bibinfo {pages} {7} (\bibinfo {year}
  {1995})}\BibitemShut {NoStop}%
\bibitem [{\citenamefont {Evans}(2007)}]{Evans:2007}%
  \BibitemOpen
  \bibfield  {author} {\bibinfo {author} {\bibfnamefont {T.}~\bibnamefont
  {Evans}},\ }\href@noop {} {\bibfield  {journal} {\bibinfo  {journal} {Eur.
  Phys.J.B}\ }\textbf {\bibinfo {volume} {56}},\ \bibinfo {pages} {65}
  (\bibinfo {year} {2007})}\BibitemShut {NoStop}%
\bibitem [{\citenamefont {Bentley}\ \emph {et~al.}(2011)\citenamefont
  {Bentley}, \citenamefont {Ormerud},\ and\ \citenamefont
  {Batty}}]{Bentley:2011}%
  \BibitemOpen
  \bibfield  {author} {\bibinfo {author} {\bibfnamefont {R.~A.}\ \bibnamefont
  {Bentley}}, \bibinfo {author} {\bibfnamefont {P.}~\bibnamefont {Ormerud}}, \
  and\ \bibinfo {author} {\bibfnamefont {M.}~\bibnamefont {Batty}},\
  }\href@noop {} {\bibfield  {journal} {\bibinfo  {journal} {Behav. Ecol.
  Sociobiol.}\ }\textbf {\bibinfo {volume} {65}},\ \bibinfo {pages} {537}
  (\bibinfo {year} {2011})}\BibitemShut {NoStop}%
\bibitem [{\citenamefont {Yule}(1925)}]{Yule:1925}%
  \BibitemOpen
  \bibfield  {author} {\bibinfo {author} {\bibfnamefont {G.}~\bibnamefont
  {Yule}},\ }\href@noop {} {\bibfield  {journal} {\bibinfo  {journal}
  {Philos.Trans.R.Soc.London B}\ }\textbf {\bibinfo {volume} {213}},\ \bibinfo
  {pages} {21} (\bibinfo {year} {1925})}\BibitemShut {NoStop}%
\bibitem [{\citenamefont {M.E.J.Newman}(2005)}]{Newman:2005}%
  \BibitemOpen
  \bibfield  {author} {\bibinfo {author} {\bibnamefont {M.E.J.Newman}},\
  }\href@noop {} {\bibfield  {journal} {\bibinfo  {journal} {Contem.Phys.}\
  }\textbf {\bibinfo {volume} {46}},\ \bibinfo {pages} {323} (\bibinfo {year}
  {2005})}\BibitemShut {NoStop}%
\bibitem [{\citenamefont {Toyokawa}\ \emph {et~al.}(2014)\citenamefont
  {Toyokawa}, \citenamefont {Kim},\ and\ \citenamefont
  {Kameda}}]{Toyokawa:2014}%
  \BibitemOpen
  \bibfield  {author} {\bibinfo {author} {\bibfnamefont {W.}~\bibnamefont
  {Toyokawa}}, \bibinfo {author} {\bibfnamefont {H.}~\bibnamefont {Kim}}, \
  and\ \bibinfo {author} {\bibfnamefont {T.}~\bibnamefont {Kameda}},\
  }\href@noop {} {\bibfield  {journal} {\bibinfo  {journal} {PLos One}\
  }\textbf {\bibinfo {volume} {9}},\ \bibinfo {pages} {e95789} (\bibinfo {year}
  {2014})}\BibitemShut {NoStop}%
\bibitem [{\citenamefont {Kameda}\ and\ \citenamefont
  {Nakanishi}(2002)}]{Kameda:2002}%
  \BibitemOpen
  \bibfield  {author} {\bibinfo {author} {\bibfnamefont {T.}~\bibnamefont
  {Kameda}}\ and\ \bibinfo {author} {\bibfnamefont {D.}~\bibnamefont
  {Nakanishi}},\ }\href@noop {} {\bibfield  {journal} {\bibinfo  {journal}
  {Evol. Hum. Behav.}\ }\textbf {\bibinfo {volume} {23}},\ \bibinfo {pages}
  {373} (\bibinfo {year} {2002})}\BibitemShut {NoStop}%
\bibitem [{\citenamefont {Yoshida}\ \emph {et~al.}(2016)\citenamefont
  {Yoshida}, \citenamefont {Hisakado},\ and\ \citenamefont
  {S.Mori}}]{Yoshida:2016}%
  \BibitemOpen
  \bibfield  {author} {\bibinfo {author} {\bibfnamefont {S.}~\bibnamefont
  {Yoshida}}, \bibinfo {author} {\bibfnamefont {M.}~\bibnamefont {Hisakado}}, \
  and\ \bibinfo {author} {\bibnamefont {S.Mori}},\ }\href@noop {} {\bibfield
  {journal} {\bibinfo  {journal} {New Generation Computing}\ }\textbf {\bibinfo
  {volume} {34}},\ \bibinfo {pages} {210} (\bibinfo {year} {2016})}\BibitemShut
  {NoStop}%
\bibitem [{\citenamefont {Rendell}\ \emph
  {et~al.}(2010{\natexlab{b}})\citenamefont {Rendell}, \citenamefont
  {Fogarty},\ and\ \citenamefont {Laland}}]{Rendell:2009}%
  \BibitemOpen
  \bibfield  {author} {\bibinfo {author} {\bibfnamefont {L.}~\bibnamefont
  {Rendell}}, \bibinfo {author} {\bibfnamefont {L.}~\bibnamefont {Fogarty}}, \
  and\ \bibinfo {author} {\bibfnamefont {K.~N.}\ \bibnamefont {Laland}},\
  }\href@noop {} {\bibfield  {journal} {\bibinfo  {journal} {Evolution}\
  }\textbf {\bibinfo {volume} {64}},\ \bibinfo {pages} {534} (\bibinfo {year}
  {2010}{\natexlab{b}})}\BibitemShut {NoStop}%
\bibitem [{\citenamefont {Bianconi}\ and\ \citenamefont
  {Barab\'{a}si}(2001)}]{Bianconi:2001-2}%
  \BibitemOpen
  \bibfield  {author} {\bibinfo {author} {\bibfnamefont {G.}~\bibnamefont
  {Bianconi}}\ and\ \bibinfo {author} {\bibfnamefont {A.-L.}\ \bibnamefont
  {Barab\'{a}si}},\ }\href@noop {} {\bibfield  {journal} {\bibinfo  {journal}
  {Phys.Rev.Lett.}\ }\textbf {\bibinfo {volume} {86}},\ \bibinfo {pages} {5632}
  (\bibinfo {year} {2001})}\BibitemShut {NoStop}%
\bibitem [{\citenamefont {Hisakado}\ \emph {et~al.}(2006)\citenamefont
  {Hisakado}, \citenamefont {Kitsukawa},\ and\ \citenamefont
  {Mori}}]{Hisakado:2006}%
  \BibitemOpen
  \bibfield  {author} {\bibinfo {author} {\bibfnamefont {M.}~\bibnamefont
  {Hisakado}}, \bibinfo {author} {\bibfnamefont {K.}~\bibnamefont {Kitsukawa}},
  \ and\ \bibinfo {author} {\bibfnamefont {S.}~\bibnamefont {Mori}},\
  }\href@noop {} {\bibfield  {journal} {\bibinfo  {journal} {J. Phys. A}\
  }\textbf {\bibinfo {volume} {39}},\ \bibinfo {pages} {15365} (\bibinfo {year}
  {2006})}\BibitemShut {NoStop}%
\end{thebibliography}%
\appendix

\section{\label{A} Exact Solvable $M=1$ case}
The Chapman--Kolmogorov equation  for
$P(n|t)=\mbox{Pr}(N_{1}(t)=n)$ is
\begin{eqnarray}
P(n|t+1)&=&\sum_{n',n''}T^{C}(n|n')T^{A}(n'|n'')\cdot P(n''|t), \nonumber \\
T^{A}(n|n')&=&q(n')\delta_{n,n'}+p_{1}(n')\delta_{n-1,n'}, \nonumber \\
T^{C}(n|n')&=&\left(1-\frac{q_{C}}{N} \right)\delta_{n,n'}+\frac{q_{C}}{N}\delta_{n,0}, \nonumber \\
p_{1}(n)&=& \frac{N-n}{N}\cdot (A\cdot (1-\delta_{n,0})+B),
\nonumber \\
q(n)&=&1-p_{1}(n).
\end{eqnarray}
The initial condition is $P(n|0)=\delta_{n,0}$.
$P(n|t)$ satisfies the following relation:
\begin{eqnarray}
P(n|t+1)&=&(1-\frac{q_{C}}{N})\cdot q(n)P(n|t)
+(1-\frac{q_{C}}{N})\cdot p_{1}(n-1)P(n-1|t)
\,\,\,\, \mbox{for} \,\,\,\, n\ge 1,  \nonumber  \\
P(0|t+1)&=&\frac{q_{C}}{N}+(1-\frac{q_{C}}{N})\cdot q(0)P(0|t).   \nonumber
\end{eqnarray}
We write the stationary probability function
as $P(n)\equiv \lim_{t\to \infty}P(n|t)$.
We solve $P(0)$ and $P(1)$ as
\begin{eqnarray}
  P(0)&=&\frac{q_{C}}{q_{C}+(N-q_{C})B},
  \nonumber \\
  P(1)&=&\frac{B}{A+B}\cdot \frac{N}{a+(N-1)}\cdot P(0) \nonumber \\
  a &=&\frac{q_{C}}{(1-q_{C}/N)(A+B)}.
\end{eqnarray}
$P(n)$ satisfies the following recursive relation for $n\ge 2$:
\[
P(n)=\frac{N-(n-1)}{N-(n-a)}\cdot P(n-1).
\]
There is a threshold value for $a$.
Depending on whether $a>1$ or $a<1$, the behavior of $p(n)$
changes drastically.
We solve the recursive relation and obtain
\begin{equation}
  P(n)=
\begin{cases}
  \frac{B(N-n+a,1-a)}{B(N-1+\alpha,1-a)}
  \cdot P(1)\,\,\,\, \mbox{for} \,\,\,\, a<1,   \\
  \frac{B(N,a-1)}{B(N-n+1,a-1)}\cdot
  P(1)  \,\,\,\, \mbox{for} \,\,\,\, a>1,   \\
  P(1) \,\,\,\,  \mbox{for} \,\,\,\, a=1.
\end{cases}
\label{eq:A3}
\end{equation}
As $B(a,b)\sim a^{-b}$,
\[
P(n)\sim (N-n)^{a-1}.
\]
We define the critical value $p_{c}$ for $p$ by
the condition $a=1$.
We obtain
\[
p_{c}=\frac{q_{C}/(1-q_{C}/N)-q_{I}}{q_{o}-q_{I}}.
\]
If $p<p_{c}$ and $a>1$, $P(n)$ becomes a
decreasing function of $n$.
However, if $p>p_{c}$ and
$a<1$, $P(n)$ becomes an increasing
function of $n$, and $P(n)$ has a fat tail.

\begin{figure}[htbp]
\begin{center}
\includegraphics[width=8cm]{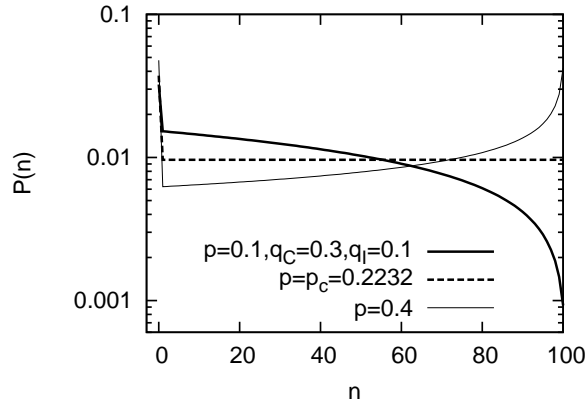}
\end{center}
\caption{\label{fig:A1}
  Plots of $P(n)$ vs. $n$ in Eq.~(\ref{eq:A3}).
  $M=1,N=10^{2},q_{c}=0.3,q_{I}=0.1,q_{o}=1$, and $p\in \{0.1,p_{c}=0.2232,0.3\}$.}
\end{figure}	

Figure \ref{fig:A1} shows a plot of $P(n)$ for
$N=10^{2},q_{c}=0.3$, $q_{I}=0.1$ and $q_{o}=1$.
In this case, $p_{c}\simeq 0.2233$. We choose $p \in
\{0.1,p_{c},0.4\}$. For $p=0.1$ and $a>1$, $P(n)$
decreases with $n$. If $p=p_{c}$ and $a=1$, $P(n)$
becomes constant for $n\ge 1$. If $p>p_{c}$, $P(n)$
has a peak at $n=N$ in addition to the peak at $n=0$.

The variance of $N_{1}/N$ is evaluated as
\begin{eqnarray}
  \mbox{V}(N_{1}/N)&\equiv& \mbox{E}((N_{1}/N)^{2})-\mbox{E}(N_{1}/N)^{2}
  \nonumber \\
&=&\frac{(2+a/N)(1+a/N)}{(a+1)(a+2)}\cdot (1-P(0))-\mbox{E}(N_{1}/N)^{2}. \label{eq:A4}
\end{eqnarray}
In the limit $N\to \infty$, we obtain
\[
\lim_{N\to \infty}\mbox{V}(N_{1}/N)=\frac{a}{(a+1)^{2}(a+2)}.
\]
As $\lim_{N\to \infty}\mbox{V}(N_{1}/N)>0$,
$N_{1}/N$ has a wide distribution in the
limit $N\to\infty$. In fact, $N_{1}/N$ obeys a beta distribution
$\mbox{beta}(1,a)$, and the correlation
coefficient among $M_{n}$
is $1/(a+2)$ in thus limit \cite{Hisakado:2006}.

\begin{figure}[htbp]
\begin{center}
\includegraphics[width=8cm]{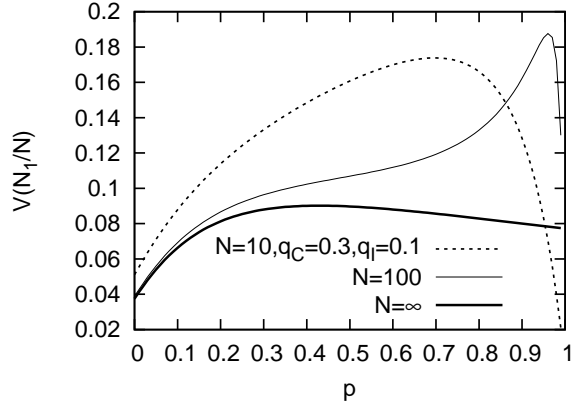}
\end{center}
\caption{\label{fig:A2}
  Plots of V$(N_{1}/N)$ vs. $p$ in Eq.~(\ref{eq:A4}).
$q_{C}=0.3,q_{I}=0.1,q_{O}=1$, and $M=1$.
$N\in \{10,10^{2},10^{4},\infty\}$.
}
\end{figure}	

Figure \ref{fig:A2} shows a plot of V$(N_{1}/N)$
vs. $p$. We adopt the same set for the parameters, as in
Figure \ref{fig:A1}. As can be clearly seen, as $N$ increases, the
curve for V$(N_{1}/N)$ converges to that for the
result in the limit of $N\to \infty$.
We find no evidence of
singular behavior in $N_{1}$.
This is an artifact of the case $M=1$.
The agents are forced to choose the unique good arm $m=1$.
With the change of the good arm,
$N_{1}$ changes to zero. As the result,
the limiting value of the variance of $N_{1}/N$ becomes
finite even for $p=0$.
If there are many good arms, agents can be distributed among multiple
good arms. If $p=0$, all agents search for a good arm
independently, which are distributed equally among all $M$ arms.
In this case, the variance of $N_{1}/N$ is proportional to $1/M^{2}$.
In the limit $p\to 1$, all agents share the same good arm, and
the variance of $N_{1}/N$ becomes finite in the limit $N\to \infty$, even
if $M\to \infty$. This means that the variance of $N_{1}$ per agent
is finite for small $p$ and diverges with $N$ for large $p$.
This suggests that the system shows a phase transition
in the limit $N,M\to \infty$.

We derive the self-consistent equation for $n_{1}$, which is the limiting value
of the expectation value of $N_{1}$, $n_{1}(t)=\mbox{E}(N_{1}(t))$.
\[
n_{1}=\lim_{t\to\infty}\mbox{E}(N_{1}(t))=\lim_{t\to\infty} \sum_{n}P(n|t)n
\]
We write the recursive relation for
$n_{1}(t)$ using the relations for $P(n|t)$ as
\[
n_{1}(t+1)=\sum_{n}P(n|t+1)\cdot n=
\left(1-\frac{q_{C}}{N}\right)
\left(1-\frac{A+B}{N}\right)n_{1}(t)+
\left(1-\frac{q_{C}}{N}\right)(A+B-AP(0|t)).
\]
We solve the self-consistent equation for
$n_{1}\equiv \lim_{t\to\infty}n_{1}(t)$
and obtain the expression in Eq.~(\ref{eq:N1_M12}).

\end{document}